\documentstyle[prl,aps,multicol]{revtex} \tighten 
\begin{document} \draft
\title{Tunneling between two semiconductors
 with localized electrons:
Can it reveal the Coulomb gap?}
\author{ A. I. Larkin and B. I. Shklovskii} 
\address{Theoretical Physics
Institute, University of Minnesota, 
116 Church St. Southeast, Minneapolis,
Minnesota 55455}
 \maketitle
 \begin{abstract}
 
It is shown that 
the voltage dependence of the tunneling conductance 
between two lightly doped semiconductors, 
which are separated by an large area tunneling 
barrier, can reveal the high energy part of 
the Coulomb gap if the barrier is thick enough.
At the barrier thickness smaller 
than average distance between impurities
no Coulomb gap feature can be found.
This happens because such tunneling is 
sensitive to very rare shortest pairs of 
occupied and empty states
localized at opposite sides of the
barrier, whose density of states in this limit 
has no Coulomb gap.
Small area tunneling contacts are also discussed.
It is shown that the tunneling conductance of a
point-like contact exponentially 
grows with the applied voltage.
This dependence does not permit
a direct measurement of the Coulomb gap.

\end{abstract}

\begin{multicols}{2}

The density of states of one-electron (or more generally charged)
excitations, $g(\varepsilon)$, 
of a three-dimensional system with localized 
electrons is known\cite{ES75,ES84} 
to have the Coulomb gap near the Fermi level, 
\begin{equation}
g(\varepsilon) = (3/\pi) \kappa^{3} e^{-6} \varepsilon^{2}~,
\label{g}
\end{equation}
where $\varepsilon$ is energy calculated from the Fermi level, $e$ 
is the proton charge
$\kappa$ is the dielectric constant of the semiconductor.
Eq.~(\ref{g}) is valid while
$\varepsilon < \Delta \sim e^2/\kappa R$,
where $R$ is the average distance between 
impurities in the semiconductor,
$\Delta$ is the width of the Coulomb gap
and simultaneously the width 
of the peak of the density of states of a lightly doped 
and moderately compensated semiconductor.

The Coulomb gap is known 
to dramatically change DC variable range hopping conductivity
leading to the Efros-Shklovskii law\cite{ES75,ES84} 
\begin{equation}
\sigma = \sigma_0 \exp[- (T_0/T)^{1/2}]~,
\label{DC}
\end{equation}
Here $T_0 = Ce^2/\kappa a$,
where $a$ is the decay length (effective Bohr radius) of impurity states and 
$C$ is the numerical coefficient.

On the other hand, the phononless high frequency
hopping conductivity is not sensitive to 
the Coulomb gap, because it is determined by very 
close pairs of empty and occupied localized states\cite{ES75,ES85}.

An appealing method of direct observation of
the Coulomb gap is low temperature tunneling 
into a semiconductor from a metallic electrode
separated by a tunneling barrier. If the barrier is thick enough,
screening of the Coulomb interaction in
the semiconductor by the metal can be neglected.
In this case, the density of localized states in the semiconductor 
can be measured studying 
variation of the tunneling conductance as a 
function of the applied voltage.
Several works\cite{Lee,Lee1,Butko}
 claim observation of the Coulomb gap 
in such a way, although, it is not totally clear why in this cases
screening by the metal was weak enough.

Recently, tunneling through the break junction in a germanium crystal 
with localized states was studied\cite{Barbara}. The tunnel 
conductance showed a minimum at zero bias voltage 
which was interpreted as the Coulomb gap. 

In this paper, we consider theoretically
tunneling between two semiconductors,
where states are localized. Both of them can be three-dimensional,
so that a barrier separates two half-spaces.
In the two-dimensional case, when electrons occupy only one plane,
the barrier can 
be a stripe separating
two half-planes filled by localized two-dimensional electrons.
Another system, with the same physics consists of  
two parallel layers of localized two-dimensional 
electrons separated by a thin barrier layer. 
For certainty, we concentrate here 
on the three-dimensional case
and assume that parameters of
semiconductors on both sides 
of the barrier are exactly the same.

Tunneling between semiconductors with discrete 
localized states has to be 
due to phonon assisted processes, 
because probability to find exactly resonant 
pairs of states on both sides of the barrier
is infinitesimally small. In this paper we 
concentrate on low temperature tunneling 
which goes with phonon emission only.

The main goal of this paper is to understand 
how the current-voltage characteristics of such a
tunneling contact depends on the barrier thickness 
$d$. If the barrier is thin, $d \ll R$, 
the distribution of electrons in the ground state
and, therefore, the Coulomb gap of the density of states
is not affected by the barrier. 
We assume that density of states 
in both semiconductors is still
given by Eq.~(\ref{g}) even if $d \gg R$\cite{Footnote}.

Below we assume that the tunneling barrier is high enough so that 
all resistance is determined by hops from 
occupied states on the left side of the barrier to empty states 
on the right side. In other words,
if we introduce the Miller-Abrahams 
resistance network\cite{ES84} 
connecting all the localized states 
inside each semiconductor and between 
them we immediately get an
enormous difference between these two types of resistors.
Resistors crossing the barrier have much 
larger resistance 
and, therefore, they determine the tunneling conductance
between two semiconductors.
One can imagine that the bulk of both semiconductors is a
superconductor and the tunneling conductance 
is due to parallel resistors which cross the barrier.
Then it is clear that at large enough $d$, the  number 
of these parallel resistors and, therefore, the tunneling 
current, $I$, is proportional to the product of number 
of localized states with energy from $0 < \varepsilon < eU/2$
on the left side and number of empty states in the interval
rom $0 > \varepsilon > -eU/2$. Each of
these numbers is proportional 
to $U^{3}$, so that the tunneling current is equal 
\begin{equation}
I = A(U)(eU)^{6}.
\label{U6}
\end{equation}
where coefficient $A(U)$ depends on $U$ through the phonon 
energy and is specified below.
We show below that at a finite barrier thickness $d$ the
current-voltage characteristic is more complicated than
Eq.~(\ref{U6}), namely, that at $U + e^2/\kappa d \leq \Delta$ 
the current is
\begin{equation}
I = A(U) eU (eU + e^2/\kappa d)^{5}.
\label{UC}
\end{equation}
In order to derive Eq.~(\ref{UC}), 
let us following Ref. \onlinecite{ES85}
introduce such a function $F(\Omega, r)$
that $F(\Omega, r) 4\pi r^{2}dr d\Omega$ is the concentration 
of pairs of occupied and empty states
with the distance between them (the arm of the pair) 
in the range between $r$ and $r + dr$
and with the energy required for electron transfer in the interval 
between $\Omega$ and $\Omega + d\Omega$. This function 
was calculated in Ref. \onlinecite{ES85} in the following way:
\begin{eqnarray}
F(\Omega, r) &=& \int_{0}^{\infty}
 \int_{0}^{\infty}  d\varepsilon_{1}d\varepsilon_{2}
g(\varepsilon_{1})g(\varepsilon_{2})
\delta(\varepsilon_{1} -
 \varepsilon_{2} - \frac{e^2}{\kappa r} - \Omega) \nonumber \\ 
&=&B\kappa^{6} e^{-12}(\Omega + e^2/\kappa r)^{5},
\label{Integral}
\end{eqnarray}
where $B =3/(10\pi^{2})$.
Using this function one can easily calculate 
the tunneling current $I$. Indeed all pairs
crossing tunneling barrier 
can contribute to $I$. They all work in parallel
because as we said above the
current reaches their ends almost without resistance.
Let us write the rate of phonon assisted tunneling hops 
across the barrier with emission of a phonon 
with energy $\Omega$ as $\nu(\Omega) = \nu_{0}(\Omega)\exp(-2d/a_0)$. 
At low enough $\Omega$
only acoustic phonons are involved. 
For emission acoustic phonons via deformational potential 
interaction $\nu_{0}(\Omega) \propto \Omega$,
when $\Omega < \hbar s/a$, 
where $s$ is the sound velocity.  
At larger $\Omega$ the rate  
$\nu(\Omega)$ sharply decreases 
with $\Omega$ because wavelength of the phonon 
becomes smaller than $a$ 
(see, for example, Ref. \onlinecite{ES84}).

Now we can identify optimal pairs which determine
the contact conductance. They all have arms which
are very close to $r=d$. Indeed, the barrier part of the arm of such
a pair can not be longer than $d + a_0$, while 
the semiconductor part of the arm 
can not be longer than $a$. Otherwise, a pair has
 exponentially smaller than optimal ones
conductance and should be disregarded. Thus, the 
occupied state of an optimal pair
should be in the layer with thickness $a$ near the plane of the left
semiconductor barrier junction. If area of the junction is 
$S$ the volume available for the occupied pair is $Sa$. 
For a given occupied state the empty state of an optimal pair
should reside in the cylinder with the height $a$ and with the 
base with the shape of the disk of the radius $(da_0)^{1/2}$
drawn on the plane of the right barrier-semiconductor junction 
around the center which exactly opposes 
the occupied state on the left junction. 
The transfer energy, $\Omega$, should be in the interval
$0 < \Omega <  eU$. Otherwise, the
pair does not consist of occupied and empty 
states and does not contribute to the tunneling current.
Thus, total number of optimal pairs 
is $N=F(eU, d)eU Sa^{2}a_{0}d$.
For the tunneling current $Ne\nu$ we arrive at 
Eq.~(\ref{UC}) where
\begin{equation}
A(U) = B\kappa^{6} e^{-11} (Sa^{2}a_{0}d)\nu_{0}(eU)\exp(-2d/a_0).
\label{A}
\end{equation} 

Looking at Eqs.~(\ref{UC}), (\ref{A}), we see that
the current-voltage characteristics 
has two voltage dependent 
factors. One of them, the rate  
of emission of phonons, $\nu_{0}(eU)$ is pretty well known.
The ratio $I(U)/U\nu(U)$ gives the energy dependence
apparent join density
of states of the two 
semiconductors.
At small $U$ this joint density of states
does not depend on $U$. 
If $d \gg R$, only at large enough voltages,
$eU > e^2/\kappa d$, one can recover the 
high energy part of the 
joint density proportional to $U^5$ and 
directly indicating existence of the Coulomb gap.
On the other hand, if $d < R$, the
joint density of states is constant 
at all $eU \leq \Delta$. At larger $U$ the tunneling 
current decreases due the decrease of the density of states
of the classical impurity band at energies larger than 
$\Delta$\cite{ES84}. Thus, at $d < R$ one 
can not see the Coulomb gap at all.

The reason for the absence of $U$-dependence 
of the joint density of states at small $d$ is in 
the strong Coulomb interaction of the electron
transferred to the right side of the barrier with the hole remaining on the left side.
The same effect leads to the enhancement of the phononless 
high frequency conductivity\cite{ES85}.
Both phenomena are determined by rare, isolated from each other
pairs of states. When comparing Eq.~(\ref{UC}) with 
formulas for the high frequency conductivity one can easily find 
straightforward analogies 
if $eU$ and $d$ of this paper are replaced by
$\hbar\omega$ and by the arm, $r_{\omega}$, of pairs  which absorb radiation quanta $\hbar\omega$. 

Our theory is applicable only to a barrier which height and thickness 
are uniform along a large area, $S$, of the contact which has many 
optimal pairs.
If area of the barrier, $S$, is so small that in average there is less than 
one optimal pair, the situation becomes mesoscopic.
This happens at $S \sim S_c$, where
\begin{equation}
S_c = \kappa^{-6}e^{12}(eU)^{-1}
(eU + e^2/\kappa d)^{-5}a^{-2}(a_{0}d)^{-1}~.
\label{Area}
\end{equation}
At $S \ll S_c$ the tunneling conductance 
is determined by the best pair which can be 
found in this area of the barrier.
Such a pair is longer than $d$ by a random length $\delta r$.
Therefore, tunneling conductance of a barrier with $S \ll S_c$
has an additional small
factor $\exp(-2\delta r/a)$. 
Of course, in such conditions the conductance 
of the tunneling contact strongly fluctuates
from a sample to sample. This is why we call this case mesoscopic.

Let us consider how growing voltage affects 
the conductance of a tunneling contact with a small
area $S$. With growing $U$ new pairs with smaller 
$\delta r$ become available  and 
the exponentially small factor, $\exp(-2\delta r/a)$,
grows. Eventually it reaches unity when 
$\delta r$ becomes as small as $a$.
Simultaneously, as we see from Eq.~(\ref{Area}),
critical area $S_c$ decreases with growing $U$
and eventually approaches $S$.
At this point we return to the conductance
of a macroscopic barrier.
This means that while $S_c \ll S$ 
the tunneling conductance
grows exponentially with voltage $U$
simulating an the joint density of states and 
the one-electron density of states $g_(\varepsilon)$
exponentially growing with energy. This phenomenon 
is, of course, different from a direct
 measurement of the Coulomb gap. 
Thus, while for the uniform barrier the Coulomb 
gap can be revealed at large enough $d$, 
the point contact does not directly reveal the Coulomb gap.

This work was supported by NSF grants DMR-0120702 and DMR-9985785.

\end{multicols}

\begin{references}
\bibitem{Footnote} Although density of states at the surface 
of a semiconductor occupying half-space is not known,
it is natural to think that it is larger than in the bulk
only by factor two or so.
The reason is that near the surface a number of limiting 
the density of states stability criteria\cite{ES75,ES84,ES85}
is twice smaller than in the bulk. If the surface 
density of states were much larger 
so that, for example, it behaved as a linear function of 
energy similarly to 
a two-dimensional system, then deeper in semiconductor 
the bulk density would be strongly 
depleted before relaxing back to the standard value.
To our mind, this is unlikely.
\end{references}
\end{document}